# Monitoring polarization in comet 46P/Wirtanen

Maxim Zheltobryukhov[1,*], Evgenij Zubko[2], Ekaterina Chornaya[3,1], Igor Luk'yanyk[4], Oleksandra V. Ivanova[5,6,4], Anton Kochergin[3,1], Gennady Kornienko[1], David Mkrtichian[7], Saran Poshyachinda[7], Igor E. Molotov[8], Sungsoo S. Kim[2,9], and Gorden Videen[10,2]

[1] Institute of Applied Astronomy of RAS, 10 Kutuzova Emb., Saint-Petersburg 191187, Russia

[2] Humanitas College, Kyung Hee University, 1732, Deogyeong-daero, Giheung-gu, Yongin-si, Gyeonggi-do 17104, South Korea

[3] Far Eastern Federal University, 8 Sukhanova St., Vladivostok 690950, Russia

[4] Astronomical Observatory, Taras Shevchenko National University of Kyiv, 3 Observatorna St., 04053, Kyiv, Ukraine

[5] Astronomical Institute of the Slovak Academy of Sciences, SK-05960 Tatranská Lomnica, Slovak Republic

[6] Main Astronomical Observatory of National Academy of Sciences, Kyiv, Ukraine

[7] National Astronomical Research Institute of Thailand, Chiang Mai 50180, Thailand

[8] Keldysh Institute of Applied Mathematics, Russian Academy of Sciences, 4 Miusskaya sq., Moscow 125047, Russia

[9] Department of Astronomy and Space Science, Kyung Hee University, 1732, Deogyeong-daero, Giheung-gu, Yongin-si, Gyeonggi-do 17104, South Korea

[10] Space Science Institute, 4750 Walnut Street, Boulder Suite 205, CO 80301, USA

**Abstract.**

We measure the degree of linear polarization of comet 46P/Wirtanen during two months, embracing the perihelion passage in December, 2018 with phase angles ranging from $\alpha = 18.1°$ to 46.4°. The polarimetric response $P_Q$ obtained resembles what was previously found in comet C/1975 V1 (West). This suggests 46P/Wirtanen belongs to a group of comets with high maximum positive polarization. In the first half of February of 2019, we conducted *BVRI* photometry of 46P/Wirtanen and found either neutral or blue colour of its dust, which is in good accordance with measurements of C/1975 V1 (West). While aperture-averaged polarimetry of 46P/Wirtanen reveals a nearly zero polarization $P_Q$ at the lowest phase angle $\alpha = 18.1°$, simultaneous imaging polarimetry suggests that the negative polarization ($P_Q < 0$) arises in a region of within 5,000 km of the nucleus, where the negative polarization could be as strong as $P_Q = -(1.44 \pm 0.15)\%$. This observation suggests the existence of the *circumnucleus halo* and that the coma is populated by at least two types of dust particles. One of those reveals a low positive polarization at side scattering and high negative polarization near backscattering. Both polarimetric features are simultaneously produced by weakly absorbing Mg-rich silicate particles. Another type of dust produces solely positive polarization that could be attributed to carbonaceous particles. This composition of 46P/Wirtanen coma appears to be similar with what was previously found in comet C/1975 V1 (West).

* Corresponding Author

E-mail address: maxim.s.zheltobryukhov@gmail.com

Phone: +7 929 428 46 26

Fax: +7 4234 39 11 21

## 1. Introduction

Comet 46P/Wirtanen was discovered by Carl A. Wirtanen on January 17, 1948. It is a Jupiter-family comet with orbital period of ~5.4 years. Comet 46P/Wirtanen is best known as a former primary target of the *Rosetta* space probe before it was rescheduled for flight to 67P/Churyumov–Gerasimenko. Despite this, the comet has not garnered much attention. For example, prior to its latest apparition in 2018–2019, the polarization of 46P/Wirtanen has not been reported in the literature. The current article is aimed at filling this gap. Using the facilities of the Ussuriysk Astrophysical Observatory, division of Institute of Applied Astronomy of RAS (code C15; Russia), we monitored the degree of linear polarization using the Johnson *V* filter over two months, between mid November of 2018 till mid January of 2019. Over this time period, observational conditions were favorable on 13 nights, providing polarimetric data at phase angles ranging from $\alpha$ = 18.1° up to 46.4°. In addition, on two epochs in the first half of February of 2019, we measured the *BVRI* photometric colour of comet 46P/Wirtanen.

Using the spectral observation of 46P/Wirtanen conducted at the Thai National Observatory of the National Astronomical Research Institute of Thailand, we estimate the relative contribution of gaseous emissions to the polarimetric measurements of 46P/Wirtanen with the *V* filter. Note, this spectral observation was aimed at identification of gaseous-emission lines in the 46P/Wirtanen coma and that will be a subject for a later publication in an independent article. Herein, we use these data solely for estimation of the ratio of flux from gaseous emissions and from continuum within the bandpass of the Johnson *V* filter. As shown in Section 3.1, this demonstrates negligible contamination from gaseous emissions.

The article is organized as follows, Section 2 briefly describes our technique of polarimetric, photometric, and spectroscopic measurements. In Section 3 we discuss results of our observations. In Subsection 3.1 we analyze the aperture-averaged polarization of comet 46P/Wirtanen versus what was previously found in other comets. In Subsection 3.2 we discuss polarimetric images of 46P/Wirtanen. Subsection 3.3 is devoted to photometric colour of its dust. Finally, in Section 4 we summarize our findings.

## 2. Observations and data reductions

### 2.1. Polarimetry

Polarimetric observations were conducted using the 22 cm telescope (*F* = 0.503 m) of the Ussuriysk Astrophysical Observatory, a division of the Institute of Applied Astronomy of RAS (code C15), which operates within the International Scientific Optical Network (ISON). The telescope was equipped with a commercially available CCD detector FLI ProLine PL4301E that has a resolution of 2084 x 2084 pixels and pixel size of 24 μm. The field of view of the CCD detector is 326 x 326 arcmin with angular resolution of 9.39 x 9.39 arcsec per pixel. The comet was imaged through the Johnson *V* filter. In addition, we use a dichroic polarization filter (analyzer) that was rotated sequentially through three fixed position angles 0°, +60°, and +120°. The exposure time was set to 30 seconds for each frame, except for the first epoch when, due to a large airmass, it was increased to 40 seconds. We repeated measurements for 12 full cycles of the analyzer rotation; only observations on January 12 of 2019 were interrupted after 10 cycles by quickly changing weather conditions. A detailed log of the observations is given in Table 1.

Three rotations of the analyzer through 60° provide data sufficient for computation of the Stokes parameters *I*, *Q*, and *U* (e.g., Chornaya et al., 2020a and references therein), fully characterizing the linear polarization in a comet. In particular, one can compute a total degree of linear polarization $P = \sqrt{Q^2 + U^2}/I$ and angle $\theta_0$ of the plane of linear polarization with respect to the first analyzer position (i.e., 0°; see, e.g., Chornaya et al., 2020a for more details). While choice of the first position could be arbitrary, in what follows, the angle $\theta_0$ can be converted to the angle between the polarization plane and the scattering plane, which we denote with $\theta$.

For the vast majority of Solar-System bodies, including comets, the angle $\theta$ takes predominantly one of two values. Namely, within the error bars, it appears equal to 0° (180°) at small phase angles $\alpha < 20–25°$ and 90° otherwise (e.g., Zubko et al., 2015). This means that the Stokes parameter $U$ determined with respect to the scattering plane nearly vanishes (e.g., Zubko and Chornaya 2019). As one can see in Table 2, this is fairly true for our observations of 46P/Wirtanen, except for $\theta \approx 44°–51°$ found on December 18 of 2018, when the comet was at phase angle $\alpha = 18.1°$. Such intermediate values of $\theta$ were earlier reported, for instance by Chernova et al. (1993) at the inversion angle, in transition from $\theta = 0°$ to $\theta = 90°$, which typically happens near $\alpha \sim 20°$. This corresponds to rotation of the polarization plane.

In planetary science, the degree of linear polarization is often considered as a sign-dependent characteristic $P_Q$ that sometimes also is denoted as $P_r$:

$$P_Q = -P\cos(2\theta) = -\frac{Q}{I} \cdot 100\% = \frac{F_\perp - F_\parallel}{F_\perp + F_\parallel} \cdot 100\%. \qquad (1)$$

Here, $F_\perp$ and $F_\parallel$ stand for fluxes of light polarized perpendicular to the scattering plane and within the scattering plane, respectively. The negative sign of $P_Q$ indicates partial linear polarization within the scattering plane ($F_\perp < F_\parallel$ or $\theta = 0°$); whereas, the positive sign indicates partial linear polarization perpendicular to the scattering plane ($F_\perp > F_\parallel$ or $\theta = 90°$).

It is finally worth noting that the 46P/Wirtanen was the third comet whose polarization we studied during the autumn of 2018. Two other comets were 21P/Giacobini-Zinner and C/2018 V1 (Machholz-Fujikawa-Iwamoto). Their detailed investigations have been reported by Chornaya et al. (2020a) and Zubko et al. (2020). All three comets were observed using the same telescope, standard stars, and data-reduction technique. We, therefore, omit their description here and refer to Chornaya et al. (2020a) for further observational details.

*2.2. Photometry*

At the beginning of February 2019, we exploited the *ChV-500* telescope ($D = 0.5$ m, $F=1.7$ m) newly installed at the Ussuriysk Astrophysical Observatory within ISON. The telescope is equipped with the *BVRI* filters from the Johnson-Cousins photometric system and a commercially available CCD detector SBIG STX-16803, having resolution of 4096 × 4096 pixels and a pixel size of 9 μm. The CCD field of view is 78 × 78 arcmin with an angular resolution of 1.14 arcsec per pixel.

Using this facility, we conducted the *BVRI* photometry of comet 46P/Wirtanen on February 8 and 10, 2019. The obtained images of the comet were processed using the *Image Reduction and*

*Analysis Facility* (IRAF). The reduction was performed with basic programs that include bias subtraction, removal of cosmic-ray events, and flat-field correction. The flat-field correction was accomplished with images of the twilight sky. Apparent magnitudes of the standard stars were adapted from the NOMAD catalogue (I/297/out in the Vizier database). It incorporates the Hipparcos, Tycho-2, UCAC2, and USNO-B1.0 et 2MASS catalogues.

We infer apparent magnitudes of 46P/Wirtanen within the projected radius of 5,000 km from the expected location of its nucleus, that, in practice, coincides with the photocenter of the comet. We captured images of the comet with short, 60-second exposures, that were repeated 10 times in each filter during the first epoch and 8 times during the second epoch. The magnitudes of 46P/Wirtanen presented in Table 1 result from averaging over all images taken with a given filter on particular epoch.

*2.3 Spectroscopy*

The spectral observations of comet 46P/Wirtanen were obtained on November 27, 2018, with the fiber-fed medium resolution echelle-spectrograph *MRES* attached to the 2.4-meter telescope at the Thai National Observatory of the National Astronomical Research Institute of Thailand (18° 34′ N, 98° 28′ E, $h$=2457 m). The spectral resolution of *MRES* is $R$=20,000 in the wavelength range 0.38 μm to 0.9 μm. Andor Newton DU940P-BV camera was used as the detector having a resolution of 2048 x 512 pixels and pixel size of 13.5 μm. The comet was near perihelion at a distance of 1.076 au from the Sun and 0.137 au from Earth with a phase angle of 46.4°. The spectrograph's input fiber cuts out about 2 arcsecs of the coma centered at the nucleus. It corresponds to the projected diameter of approximately 200 km. During the night, three spectra were obtained with an exposure 2400 sec for of each of them. Observational data have been processed using a flexible pipeline echelle-spectra reduction code *HiFLEx* (available at https://github.com/ronnyerrmann/exohspec). The code was developed by Errmann et al. (*in preparation*). Note, the spectral observations of 46P/Wirtanen are not the subject of this paper. We use them solely for estimation of the ratio of flux from gaseous emissions and from the continuum within the bandpass of the Johnson *V* filter. The spectral observations, data reduction, and analysis focused on identification of the emission lines in 46P/Wirtanen and will be fully described in a separate article.

**3. Results and discussion**

*3.1. Aperture-averaged degree of linear polarization in comet 46P/Wirtanen*

We analyze the aperture-averaged polarization in comet 46P/Wirtanen measured with two aperture radii of ~5000 km and ~10,000 km. Considering a small-sized aperture provides two important advantages, which we describe below. The first pertains to gas emission. We study polarization of comet 46P/Wirtanen using the Johnson *V* filter whose bandpass transmits not only the *continuum* caused by elastic light scattering from cometary dust, but also emissions from cometary gases (e.g., Farnham et al., 2000; Ivanova et al., 2013). The strongest contamination in the *V* filter could arise from the $C_2$-molecule emission in the Swan band near $\lambda$ = 0.5165 μm, which might potentially depolarize the light scattered from the continuum (e.g., Le Borgne et al., 1987; Sen et al., 1989).

It is important, however, that the $C_2$ molecules within the coma result from photodissociation of their parent molecules (e.g., Jackson et al., 1991) and/or carbonaceous dust particles (e.g., Clairemidi et al., 1990; Rousselot et al., 1994). This process takes a relatively long time, so the $C_2$ molecules appear in considerable quantities only at a significant distance from the nucleus. In the *Haser model*, this characteristic is referred to as the *scale length* of the $C_2$ parent molecules (e.g., Combi et al., 2004). Several empirical estimations of the scale length of the $C_2$ parents have been reported in the literature. We note that this estimate tended to grow over the years. While A'Hearn (1982) inferred only 17,000 km at a heliocentric distance $r_h$ = 1 au, Cochran (1985) found 25,000 km, Krasnopol'skii et al. (1986) suggested 30,000 km; whereas, Fink et al. (1991) increased the distance to 58,000 km. The $C_2$ parent-molecule scale length increases with the heliocentric distance as $r_h^2$ (Fink et al., 1991) or even faster, as $r_h^{2.5}$ (Cochran 1985). This suggests that the $C_2$ emission lines fully disappear from the 46P/Wirtanen spectrum at $r_h \geq 2.4$ au (Schulz et al., 1998). However, at given $r_h$, the risk of $C_2$ emission contamination can be substantially reduced by decreasing the aperture. For instance, Clairemidi et al. (1990) investigated the *in situ* ratio of fluxes from dust and gaseous components of the 1P/Halley coma when it was at $r_h$ = 0.83 au. It was clearly demonstrated that the flux integrated along the line of sight that was shifted 3,125 km from the nucleus location was predominantly governed by the continuum. Gaseous emission is comparable to the continuum when the line of sight is shifted to 34,504 km. The direct integration of the spectrum measured in the innermost coma suggests only < 3% contribution of the gaseous emission into the flux in the visible (e.g., Jewitt et al., 1982; Picazzio et al., 2019). In other comets, it may grow up to 10% (Kwon et al., 2017). Using the approach of Jewitt et al. (1982) and Picazzio et al. (2019), we investigate the 46P/Wirtanen spectrum obtained as described in Section 2.3 and find the gaseous-emission contribution into the *V*-filter flux to be only ~5.4%.

$C_2$ molecules emit low-polarized radiation. The maximum of their linear polarization $P_{max}$ = 7.7% is attained at α = 90° (e.g., Le Borgne et al., 1987; Sen et al., 1989); whereas, at other phase angles, it takes on even smaller values. For instance, at α = 46.4°, as was in Comet 46P/Wirtanen on November 27, 2018, the degree of linear polarization of the $C_2$ emission computed with Eq. (1) of Le Borgne et al. (1987) is as low as $P_Q$ = 3.9%. Taking into account that, on the same epoch, the $C_2$ emission within the *V* filter does not exceed 5.4%, one can constrain its maximum impact on the polarization of Comet 46P/Wirtanen to be less than 0.21%. However, within the 5000-km aperture, we found $P_Q$ = (6.92 ± 1.15)%. Therefore, the $C_2$-emission impact is a factor of 33 smaller than the polarimetric response measured in Comet 46P/Wirtanen. It also is about five times smaller than the uncertainty in our polarimetric measurement.

The second advantage of the small aperture is that the inner coma (cometocentric distance of ~5,000 km) is dominated by dust particles whose age is of 1–3 days (Zubko et al., 2015; Luk'yanyk et al., 2019; Kochergin et al. 2019). This simultaneously implies that over this time period, the inner-coma population will be fully replenished. Thus, the small aperture is best suited for searching inhomogeneity in emanations of dust from a cometary nucleus. If those exist, this could be observed through fast variations of the colour and/or linear polarization of innermost coma (Lara et al., 2003; Ivanova et al., 2017; Luk'yanyk et al., 2019; Chornaya et al., 2020a).

In Fig. 1 we show images of 46P/Wirtanen as it appeared on four epochs embracing our observational campaign. The red point (see online version for reference to the colour) here shows a photometric center that also corresponds to the nucleus location. Two circles demonstrate areas of integration with radii of 5,000 km (yellow) and 10,000 km (red). Table 2 presents results for $P$, $\theta$, and $P_Q$, with their corresponding error bars. As one can see, an increase of the aperture radius from 5,000 km to 10,000 km hardly affects the polarimetric response. Polarization tends to grow slightly in the larger aperture, which is a somewhat ambiguous conclusion due to the error bars.

It is worth noting that due to low spatial resolution in the beginning and end of our observations of 46P/Wirtanen, ~1000–1500 km/px, a spatial radius of 5000 km is equivalent to only 3.4–5 pixels, depending on the epoch. Such low resolution may cause a risk of misidentification of the photometric center. Such an error in theory may affect the results because the fluxes integrated at three different orientations of the analyzer would correspond to somewhat displaced areas in the 46P/Wirtanen coma. In practice, however, we did not face ambiguity in detecting the photometric center as, on average, it was ~20% brighter compared to the second brightest pixel. However, we simulate a hypothetical misidentification of the photometric center in order to investigate its possible effect on the resulting values of the degree of linear polarization. For instance, on two epochs, November 16 and December 1, in one image out of three obtained for each full cycle of the analyzer rotation we artificially assigned the second brightest pixel to be the photocenter of the 46P/Wirtanen coma. This causes a shift of the aperture as compared to the case of correct identification. However, the displaced aperture still embraced the true photocenter as it always appears to be a neighbor to the second brightest pixel. For November 16, such an offset yields $P = 5.24\%$ versus a value of $P = 5.43\%$ when the brightest pixel is used; whereas, for December 1, we obtained $P = 6.29\%$ and 6.06%, respectively. Clearly, in both cases, the net effect of misidentification of the photometric center is a factor of 5 less than the already existing uncertainty in our polarimetric measurements (see Table 2). The reason for this is the very rapid drop in brightness in the inner coma with growing cometocentric distance.

Interestingly, during the first fifteen days of observations, the phase angle was fairly constant $\alpha = (45.5 \pm 0.9)°$. Over this time period, the degree of linear polarization did not reveal noticeable variations. During the first epoch, we observed 46P/Wirtanen at quite large airmass, 4.677 (see Table 1). In the next two weeks, it dropped to 2.492. Despite this significant change, we detect nearly the same polarimetric response. Such stability of the polarimetric response demonstrates the high performance of the correction for the airmass that we apply to all images of 46P/Wirtanen.

The other four observations, one made prior to the perihelion passage and the subsequent three, can be grouped around phase angle $\alpha = (33.0 \pm 0.4)°$. They agree with each other within the error bars. When plotting all the observations together, see blue points in Fig. 2, one can see that 11 out of 13 points follow the same trend; whereas, the values obtained at $\alpha = 28.2°$ and $23.8°$ noticeably deviate from that trend. Those two observations closely correspond to the perihelion passage on December 12 of 2018 (see Table 2).

These observations suggest high stability of the microphysical properties of dust in the 46P/Wirtanen coma. We found changes only over a short, 2-day time period embracing the perihelion passage. Taking into account that the dust particles cannot settle in the inner coma for extended periods (Zubko et al., 2015; Luk'yanyk et al., 2019), this suggests a highly homogenized emanation of dust from the 46P/Wirtanen nucleus. The radius of the 46P/Wirtanen nucleus was estimated to be (555 ± 40) m under an assumption of the geometric albedo $A = 0.04$ (Boehnhardt et al., 2002). Such an estimation should be considered as an upper limit because the surface reflectivity could be much larger. For instance, *in situ* studies of the 67P/Churyumov-Gerasimenko nucleus has revealed $A = 0.062 \pm 0.002$ (Ciarniello et al. 2015). Interestingly, even under the assumption of a 600 m radius of the 46P/Wirtanen nucleus, its production rate of water molecules suggests about 100% of its surface is active (Fink et al. 1998). Therefore, it seems unlikely that such a small and active body might have a significant compositional inhomogeneity. One could hypothesize that a short-term increase of the linear polarization is caused by an outburst from 46P/Wirtanen. The increase of brightness on December 12 of 2018 supports this hypothesis (see magnitudes in Table 1). Moreover, outburst activity of 46P/Wirtanen on its 2018 apparition has been reported by Kelley et al. (2019). While such a change in polarization can indicate the release of materials of new composition, it is worth noting that it is also can indicate a change in the volume ratios of materials already present in the coma. As demonstrated below, we can model the 46P/Wirtanen coma using two types of dust particles; e.g., one having Mg-rich silicate composition and another having carbonaceous composition. These two components reveal noticeably different polarimetric responses. Changes in their relative abundance may increase polarization of 46P/Wirtanen during its outburst activity.

The smallest phase angle achieved in our polarimetric observations of comet 46P/Wirtanen was $\alpha = 18.1°$. At such α, the degree of linear polarization $P_Q$ changes its sign that is representative of changing the polarization-plane orientation (e.g., Chernova et al., 1993). Within the 5,000-km aperture, we obtain $P_Q = (0.28 \pm 0.47)\%$; whereas, $P = (3.06 \pm 0.40)\%$ and $\theta = -(47.66 \pm 3.75)°$. Clearly, $P_Q$ acquires its small value not due to small $P$, but because the cosine function of the doubled θ approaches zero. In other words, the degree of linear polarization $P$ is primarily governed by the Stokes parameter $U$ instead of $Q$. This feature has an extremely important implication for coma composition. First of all, it reveals a minor contribution of the $C_2$ emission as its linear-polarization plane is always perpendicular to the scattering plane, i.e., $\theta = 90°$ (e.g., Le Borgne et al., 1987). Thus, unlike the continuum, the gaseous emission does not produce negative linear polarization.

While particle alignment can produce non-zero $U$, it should be ruled out for producing non-zero $U$ at $\alpha = 18.1°$ because it would produce a similar non-zero value at other phase angles. However, this contradicts our measurements of 46P (see Table 2) and other comets (e.g., Chornaya et al. 2020a; Zubko et al. 2020). As one can see, $P$ is solely governed by the $Q$ parameter, suggesting a vanishingly small contribution of the $U$ component. This feature is consistent with observations of comets reported by other groups (e.g., Chernova et al. 1993; Levasseur-Regourd et al. 1996). Moreover, an alignment should yield $P \neq 0$ at the exact backscattering, $\alpha = 0°$ (Bohren and Huffman 1983); however, this also has not been observed as α approaches 0° (e.g., Chernova et al. 1993; Levasseur-Regourd et al. 1996).

Rotation of the linear-polarization plane is inconsistent with the scattering of unpolarized sunlight from randomly oriented irregularly shaped particles. Indeed, their degree of linear polarization is solely governed by the Stokes parameter $Q$: $U = 0$, $P_Q = -P\cos(2\theta)$ (Zubko and Chornaya 2019). As a consequence, a change of the polarization sign is caused by a change of the sign of $Q$. This occurs at the inversion phase $\alpha_{inv}$ where $Q = 0$. At this point $P = P_Q = 0$; whereas, in comet 46P/Wirtanen we detect $P \neq 0$ and $P_Q = 0$, well above the measurement errors.

One possible explanation of this paradox is that the 46P/Wirtanen coma is comprised of dust particles having different scattering properties located in different regions of the coma, and that the field aperture covers a finite range of phase angles. To explain this further, consider the coma to consist of two different types of dust. The Stokes parameter $Q_{mix}$ of their mixture is defined then as $Q_{mix} = Q_1 + Q_2$, where the subscripts correspond to dust of type 1 and 2. Evidently, $Q_{mix} = 0$ no longer requires simultaneous $Q_1 = 0$ and $Q_2 = 0$. Instead, $Q_{mix} = 0$ may result from $Q_1 = -Q_2$; i.e., one component of the mixture must produce negative polarization. This conforms with polarimetric observations of comets at the inversion angle. For instance, dust particles can consist of highly absorbing particles, whose linear polarization in the backscatter region is entirely positive and nonabsorbing particles that have negative polarization in the backscatter region. Such absorbing particles are seen in jets and such nonabsorbing particles are seen in the circumnucleus halo, both of which have been seen simultaneously at small phase angles (see, e.g., review by Hadamcik and Levasseur-Regourd 2003a; and analysis in Zubko et al. 2012).

To demonstrate how such a system can explain this paradox, let us consider, for simplicity, two single-scattering dust particles with their four-dimensional Stokes vectors (e.g., Bohren and Huffman 1983) expressed in a form:

$$S_1 = \begin{pmatrix} I_1 \\ Q_1 \\ 0 \\ 0 \end{pmatrix} \text{ and } S_2 = \begin{pmatrix} I_2 \\ Q_2 \\ 0 \\ 0 \end{pmatrix}. \tag{2}$$

We also assume that at a given phase angle, $\alpha_{inv}$, both particles produce a non-zero linear polarization with $Q_1 \neq 0$ and $Q_2 \neq 0$, whose sum is zero, i.e., $Q_1 + Q_2 = 0$. It is important that the Stokes vector is defined exclusively with respect to some reference plane, which in practice is set to be the scattering plane. If both test particles correspond to the same scattering plane, their cumulative Stokes vector can be determined by the sum of Stokes vectors of the individual particles. This would yield unpolarized light with the Stokes vector

$$S = S_1 + S_2 = \begin{pmatrix} I_1 + I_2 \\ 0 \\ 0 \\ 0 \end{pmatrix}. \tag{3}$$

In practice, the field aperture is not vanishingly small, and scatterers at each location have a slightly different phase angle and a reside within a local scattering plane, defined by the source, particle and detector. Depending on the particles' locations, their local scattering planes may not be parallel to the reference scattering plane, nor to each other's scattering plane. In order to calculate the measured Stokes vector from both particles, their Stokes vectors cannot be immediately added to one another as it was done in Eq. (3). They first must be rotated to be parallel with the reference scattering plane. This can be done via rotation of the particles' Stokes vectors (see, e.g., Eq. (2.83) of Bohren and Huffman 1983):

$$S'_j = \begin{pmatrix} 1 & 0 & 0 & 0 \\ 0 & \cos(2\psi_j) & \sin(2\psi_j) & 0 \\ 0 & -\sin(2\psi_j) & \cos(2\psi_j) & 0 \\ 0 & 0 & 0 & 1 \end{pmatrix} S_j. \qquad (4)$$

where $j$ denotes the particle and the angle $\psi_j$ is measured between the scattering plane of particle $j$ and the reference plane. The measured Stokes vector $S$ is the superposition of the Stokes vectors from both particles and is expressed as follows:

$$S = S'_1 + S'_2 = \begin{pmatrix} I_1 + I_2 \\ Q_1 \cos(2\psi_1) + Q_2 \cos(2\psi_2) \\ -Q_1 \sin(2\psi_1) - Q_2 \sin(2\psi_2) \\ 0 \end{pmatrix}. \qquad (5)$$

It is significant that in the ground-based observations, a coma appears at a quite small angular size, well below 1 arcmin. This size roughly corresponds to the upper limit in variation of the angle $\psi$ throughout the coma. As $y \to 0$, $\cos(y)$ and $\sin(y)$ can be approximated with $(1 - y^2)$ and $y$, respectively, where the angle $y$ is measured in radians. Since $y \gg y^2$, one can further simplify the approximation of the cosine function to $\cos(y) \approx 1$. Upon substitution, the Stokes vector further reduces to

$$S = S'_1 + S'_2 = \begin{pmatrix} I_1 + I_2 \\ Q_1 + Q_2 \\ -2\psi_1 Q_1 - 2\psi_2 Q_2 \\ 0 \end{pmatrix} \begin{pmatrix} I_1 + I_2 \\ 0 \\ -2\psi_1 Q_1 - 2\psi_2 Q_2 \\ 0 \end{pmatrix}. \qquad (6)$$

In this example, the measured Stokes parameter $U$ vanishes when a multiple-particle system appears to have $\psi_1 = \psi_2$. However, if $\psi_1$ and $\psi_2$ have opposite sign, e.g., $\psi_1 = -\psi_2$, such asymmetry in the distribution of dust particles with regard to the reference plane compensates an opposite sign in $Q_1$ and $Q_2$, yielding $U \neq 0$ simultaneously with $Q = 0$. This could happen, for instance, if one particle is above the reference scattering plane and another particle is below the reference scattering plane. In a cometary coma, this could happen if a jet distributes absorbing particles heterogeneously. Such heterogeneous distributions of polarization have previously been observed in numerous comets (e.g., Hadamcik and Levasseur-Regourd 2003a; Hadamcik et al. 2014; Hines et al. 2014; Chornaya et al. 2020a). This is what we find in comet 46P/Wirtanen on December 18 of 2018, where we do find a heterogeneous distribution in the polarimetric image of comet 46P/Wirtanen (see the next Section). Thus, the rotation of the linear-polarization plane at the inversion phase angle could be considered as an indicator of heterogeneous distributions of polarization in a coma. Unfortunately, a quantitative interrelation between the polarization strength at the inversion point and asymmetry in the distribution of polarization within the coma is seemingly inaccessible in a general form. This is because the amplitude of linear polarization is the result of a superposition of the signals from a 3D spatial distribution of dust particles of different types within the cometary coma; whereas, the image is a 2D projection of the heterogeneous distribution of the dust population. Unfortunately, the retrieval of the distribution of dust particles from such limited data is an ill-posed problem.

In Fig. 2, we plot the degree of linear polarization $P_Q$ as a function of phase angle α in comet 46P/Wirtanen versus what was measured with the broadband $V$ filter in other comets; C/1975 V1 (West) as reported by Kiselev and Chernova (1978), 21P/Giacobini-Zinner measured in the 1985 apparition by Kurchakov et al. (1986) and in the 2018 apparition by Chornaya et al. (2020a), and C/2018 V1 (Machholz–Fujikawa–Iwamoto) adapted from Zubko et al. (2020). Clearly, polarization of dust in comet 46P/Wirtanen is considerably lower as compared to 21P/Giacobini-Zinner and it better corresponds with the case of C/2018 V1 (Machholz–Fujikawa–Iwamoto). This might suggest that 46P/Wirtanen could belong to a class of comets with low maximum of positive polarization $P_{max}$ (e.g., Chernova et al., 1993; Levasseur-Regourd et al., 1996; Zubko et al., 2016).

However, over the available range of α, the polarization of 46P/Wirtanen nearly coincides with what was found in C/1975 V1 (West) in a similar broadband $V$ filter. Although polarimetric measurements of comet West were limited in the $V$ filter to relatively small phase angles, α < 52°, its polarization was also studied with the green continuum filter ($\lambda_{eff}$ = 0.53 µm, $\Delta\lambda$ = 0.005 µm) over a much wider range of α, from 14° up to 98°. That extended set of data suggests $P_{max}$ ≈ 22% in comet C/1975 V1 (West), which is consistent with the high-$P_{max}$ comets (Zubko et al., 2014). We do note, however, that the linear polarization of comet West in the $V$ filter and in the green continuum filter do not match perfectly.

It is significant that the microphysical properties of dust in comet C/1975 V1 (West) have been retrieved by Zubko et al. (2014) through a comprehensive model of its phase function in the visible, angular profile of polarization in the green continuum, photometric and polarimetric colours. The modeling suggested that the West coma is populated by a mixture of weakly absorbing Mg-rich silicate particles (~1/4 by volume) and highly absorbing amorphous-carbon particles (~3/4 by volume). The resemblance of the polarimetric response in comets 46P/Wirtanen and C/1975 V1 (West) in the broadband $V$ filter suggests similar microphysics of their dust and, thus, the findings in Zubko et al. (2014) could be extrapolated for the case of 46P/Wirtanen. Such a two-component composition of the 46P/Wirtanen coma also would be consistent with the polarization-plane rotation in 46P/Wirtanen at α=18.1° that was discussed above.

Finally, an increase of the polarization on December 12 and 14 of 2018, at α = 28.2° and 23.8°, respectively, could be explained as a temporal increase of the concentration of the highly absorbing particles in the innermost coma of 46P/Wirtanen. Such particles produce a high positive polarization at large α and no negative polarization at small α (Zubko et al., 2012; 2013; 2014). Interestingly, their large abundance in the coma on December 18 of 2018 could also explain a nearly zero polarization $P_Q$ of 46P/Wirtanen at α=18.1°. Therefore, the larger concentration of highly absorbing, carbonaceous particles that appeared in 46P/Wirtanen on its perihelion passage could last for a week or even longer. Our next observation was conducted only two weeks later, and it revealed polarization consistent with the pre-perihelion period. We note that the effect of the solar-radiation pressure is considerably greater for carbonaceous particles compared to the silicate particles (Zubko et al., 2015). This implies that the carbonaceous particles get swept out of the inner coma faster. As a consequence, to keep their large concentration in the inner coma, one needs to suggest a long-lasting ejection process; during, at least, a week.

*3.2. Polarimetric images of comet 46P/Wirtanen*

For six epochs, between December 4 of 2018 and January 12 of 2019, we obtained polarimetric images of the 46P/Wirtanen coma which are presented in Fig. 3. All these images look very much similar to one another. Namely, the innermost coma (radius of ~5,000–10,000 km) reveals a systematically lower positive linear polarization. The spatial distribution of the polarization is highly symmetric in appearance except for December 18, 2018. Moreover, on that epoch we detect an extended area of the negative polarization. While the aperture with radius of 5,000 km comprises 277 pixels, the area of negative polarization spans 66 pixels, i.e., about 1/4 of the entire aperture. The strongest negative polarization that we found in the 46P/Wirtanen coma was $P_Q = -1.59\%$. Averaging of this pixel and its three closest neighbors with negative polarization yields $P_Q = -(1.44 \pm 0.14)\%$. As explained in the previous section, a non-symmetric distribution of the polarization with the simultaneous presence of positive and negative polarization is consistent with the rotation of the polarization plane found at $\alpha = 18.1°$.

Fig. 4 demonstrates an asymmetry in spatial distribution of the polarization defined through the Stokes parameter *U*, $P_U = -U/I = -P\sin(2\theta)$, and the angle $\theta$ in the 46P/Wirtanen coma on December 18, 2018. As noted in the previous section such asymmetry may tentatively explain the phenomenon of rotation of the linear-polarization plane observed in various comets at the inversion phase angle.

Polarization $P_Q = -(1.44 \pm 0.14)\%$ detected in the 46P/Wirtanen coma significantly exceeds what is typically seen in a whole comet at a similar phase angle (see, e.g., Chernova et al., 1993; Levasseur-Regourd et al., 1996). However, imaging polarimetry of some comets at small phase angles does reveal a very strong negative polarization in the vicinity of their nuclei (Hadamcik and Levasseur-Regourd 2003a; 2003b). This feature is referred to as the *circumnucleus halo*. Therefore, our observations suggest the presence of such halo in comet 46P/Wirtanen.

It is significant that the circumnucleus halo is characterized not solely by its strong negative polarization at small phase angles, but, simultaneously, by a weak positive polarization at side scattering (Hadamcik and Levasseur-Regourd 2003a; 2003b). The latter feature is consistent with *in situ* polarimetric measurements of comets 1P/Halley and 26P/Grigg–Skjellerup by the *Giotto* space probe, that revealed a significant decrease of positive polarization in the inner coma (e.g., Levasseur-Regourd et al., 2005). The circumnucleus halo reveals typically a small extent as compared to other features in the coma. This was reasoned by small ejection velocity of the dust particles forming the halo (Zubko et al., 2012).

Zubko et al. (2012; 2013) have demonstrated that the circumnucleus halo is formed by dust particles with low material absorption. Imaginary part of their complex refractive index is constrained to $\text{Im}(m) \leq 0.02$ in the visible, which is consistent with Mg-rich silicates (e.g., Dorschner et al., 1995). Such particles have been found *in situ* in comets (e.g., Fomenkova et al., 1992; Ishii et al., 2008). Mg-rich silicates have been unambiguously identified with the so-called *10-μm silicate feature* often observed in the mid-IR spectra of comets (e.g., Gehrz and Ney 1992; Hanner and Bradley 2004; Wooden et al., 2017; Chornaya et al. 2020b). The presence of weakly absorbing particles also is required to fit the aperture-averaged polarimetric observations of numerous comets (Zubko et al., 2014; 2016; 2020; Chornaya et al., 2020a), including comet C/1975 V1 (West) whose polarimetric response in the *V* filter resembles what we found in 46P/Wirtanen (see Fig. 1).

Comet 46P/Wirtanen reveals a systematically lower positive polarization in its inner coma on all epochs presented in Fig. 3. This observation supports the presence of the circumnucleus halo.

However, it is worth noting that the halo could be hidden by dusty jets in comets (Hadamcik and Levasseur-Regourd, 2003a). Unlike the circumnucleus halo, dusty jets and arcs produce no negative polarization at small phase angles; whereas, at large phase angles their positive polarization attains values considerably larger than in the whole comet (e.g., Hadamcik and Levasseur-Regourd, 2003a; 2003b; Hines et al., 2014). The polarimetric response of dusty jets in comets is consistent with the domination of highly absorbing carbonaceous particles (e.g., Zubko et al., 2013; 2015). Note that micron-sized particles with such chemical composition have been indeed found in comets *in situ* (e.g., Fomenkova et al., 1992).

Although no jet-like structures with a strong positive polarization appear on the maps in Fig. 3, large positive polarization in the outer coma is overall consistent with the presence of dusty jet activity. It could happen that we cannot discriminate individual dusty jets because of the large active surface on the 46P/Wirtanen nucleus. Numerous jets evenly distributed over the nucleus may produce an impression of the homogenized outflow of cometary dust to the observer. However, this would be inconsistent with the negative polarization at α = 18.1° in the innermost coma and low positive polarization at larger phase angles. It would be a more plausible suggestion that we observed a superposition of the halo and numerous dusty jets.

We search for minimum values of the degree of linear polarization $P_Q$ in the maps on Fig. 3 and plot them in Fig. 5 versus what was found in the circumnucleus halo of comets C/1990 K1 (Levy), C/1995 O1 (Hale-Bopp), 22P/Kopff, 47P/Ashbrook-Jackson, and 81P/Wild 2; the data adapted from Hadamcik and Levasseur-Regourd (2003a; 2003b). As one can see here, at α < 30°, 46P/Wirtanen reveals somewhat weaker negative polarization and considerably stronger positive polarization than the circumnucleus halo in other comets; whereas, at α > 30°, they are much more consistent with one another. This suggests that the circumnucleus halo was partially hidden by jet activity in 46P/Wirtanen on its perihelion passage and, to some extent, one week after that Interestingly, the minimum value of positive polarization in the inner coma of 46P/Wirtanen detected on December 10 of 2018 at α = 32.6° appears in accordance with that on January 17 of 2019, at α = 33.4°. This would be consistent with a similarly small level of jet activity on both epochs.

### 3.3. Colour of the inner coma in comet 46P/Wirtanen

Similar to the degree of linear polarization, photometric colour is independent of the number of dust particles in a cometary coma and, therefore, it immediately addresses their microphysical properties. We infer three colour indices $\Delta m$ = B–V, V–R, and R–I in the innermost coma of 46P/Wirtanen (aperture radius of ~5,000 km) that have been compensated for the solar-radiation colour in the Johnson-Cousins photometric system (Holmberg et al., 2006). The results are presented in Table 3. However, in the literature, photometric colour of cometary dust is often characterized with the colour slope $S'$ that can be inferred from the colour index $\Delta m$ as follows:

$$S' = \frac{10^{0.4\Delta m} - 1}{10^{0.4\Delta m} + 1} \times \frac{20}{\lambda_2 - \lambda_1}. \tag{7}$$

Here, $\lambda_1$ and $\lambda_2$ denote the effective wavelengths of the filters. They are measured in μm and selected such that $\lambda_2 > \lambda_1$. The colour slope $S'$ is measured in % per 0.1 μm. Similar to $\Delta m$, $S' < 0$ indicates blue colour and $S' > 0$ indicates red colour.

Using Eq. (7) and effective wavelengths of the *BVRI* filters $\lambda_B = 0.433$ μm, $\lambda_V = 0.55$ μm, $\lambda_R = 0.64$ μm, and $\lambda_I = 0.79$ μm, we compute the colour slope $S'$ whose values also are presented in Table 3. What immediately emerges from these data is that on February 8, 2019, dust in comet 46P/Wirtanen had a neutral colour in the pair of *B* and *V* filters, and certainly a blue colour in two other pairs of filters. Three days later, on February 10, the same part of the 46P/Wirtanen coma was neutral in all three pairs of filters.

It is significant that similar neutral and blue colour also was noticed in comet West (see Zubko et al., 2014). However, our two photometric observations may be insufficient to fully characterize the colour of the dust in 46P/Wirtanen. Indeed, it was demonstrated in other comets that colour can be subject to fast and dramatic variations in comets (Lara et al., 2003; Weiler et al., 2003; Ivanova et al., 2017; Luk'yanyk et al., 2019). Therefore, it is of high practical interest to compare results of our photometric observations with what was reported by other teams for 46P/Wirtanen. It is worth noting, however, that the photometric colour of micron-sized dust particles is a function of phase angle α (Zubko et al., 2014), which can make such comparisons somewhat difficult.

We note that the dust of 46P/Wirtanen appears somewhat bluer than that of Comet 67P/Churyumov-Gerasimenko at a small phase angles. Using the Bessel *R* and *I* filters, Sen et al. (2019) reported $S' = (21.5 \pm 6)$ % per 0.1 μm in Comet 67P/Churyumov-Gerasimenko; whereas, with a similar pair of filters we found a colour slope of 46P/Wirtanen to be either about zero or slightly negative. We did use a somewhat smaller aperture, 5000 km versus 11,000 km of Sen et al. (2019). It also is worth noting that colour imaging revealed a decrease of reddening in the inner coma of 67P/Churyumov-Gerasimenko (see Fig. 7 of Sen et al., 2019).

For the 2018 apparition, Kelley et al. (2019) reported the colour index $g - r = (0.45 \pm 0.02)$ mag of comet 46P/Wirtanen in the *PanSTARRS 1 photometric system*. This colour index was not compensated for the solar radiation, 0.39 mag (Kelley, *private communication*). Thus, the comet 46P/Wirtanen was only slightly red in appearance, $g - r = (0.06 \pm 0.02)$, in the second half of September, 2018. This result is consistent with our own finding. Kelley et al. (2019) observed comet 46P/Wirtanen at α ≈ 23°, which is quite close to our observations at α ≈ 27°. Moreover, Kelley et al. (2019) considered the very innermost coma with an aperture radius of ~3,000 km, which is close to our 5,000 km. The time period between these two observations is approximately 5.5 month, and it would be unreasonable to expect an exact match of the colour of 46P/Wirtanen on both epochs.

Interestingly, Fink et al. (1998) reported a moderately red colour of the continuum in 46P/Wirtanen on its 1997 apparition. This conclusion was drawn from the fit of high-resolution spectra of the comet. Schulz et al. (1998) also inferred continuum spectra of 46P/Wirtanen from the same apparition. Although they did not discuss the continuum colour, their Fig. 1 reveals all three options, neutral, red, and blue colour of dust over the range of wavelength λ = 0.5 – 0.6 μm. About a half year prior to the 1997 perihelion passage, when the comet was at $r_h = 2.99 - 1.83$ au, Meech et al. (1997) measured the coma colour using the broadband *BVRI* filters. They found that the coma plus nucleus were slightly redder than the Sun. At the same heliocenetric distance, but ~5.5 year later, Boehnhardt et al. (2002) detected an extremely red colour of the inner coma and nucleus in 46P/Wirtanen, $S' = 47$% per 0.1 μm.

$S'$ reported in Boehnhardt et al. (2002), to our knowledge, differs dramatically from all other results of photometric observations of the 46P/Wirtanen coma. Moreover, such colour slope also seems to be outstanding for other comets. It is worth noting that the colour slope of the 46P/Wirtanen nucleus was estimated to be $S' = 10$% per 0.1 μm (Lamy et al., 1998) that makes it impossible to explain the finding by Boehnhardt et al. (2002) by the contribution of the nucleus. We can only suggest some peculiar activity in the inner coma of 46P/Wirtanen during its observations by Boehnhardt et al. (2002). As shown in Zubko et al. (2015), $S' = 47\%$ per 0.1 μm unambiguously suggests a relatively strong wavelength dependence of the imaginary part of refractive index Im($m$), its difference at the effective wavelengths of two filters should be not less than 0.03; whereas, index $n$ in the power-law size distribution of cometary dust $r^{-n}$ takes on relatively small values, $n < 2.5$.

## 4. Conclusion remarks

We measure the degree of linear polarization in comet 46P/Wirtanen using the broadband Johnson *V* filter during two months embracing its perihelion passage in December of 2018. The comet was available for observations at phase angle ranging from α = 18.1° to 46.4°, which makes it somewhat difficult to infer the negative-polarization and positive-polarization branches in 46P/Wirtanen. Although the aperture-averaged polarization was very nearly zero at the lowest phase angle, polarimetric image obtained on that epoch reveals negative polarization in the vicinity of the 46P/Wirtanen nucleus, up to $P_Q = -(1.44 \pm 0.15)\%$. Moreover, about 1/4 of the coma projected area within 5,000 km from the nucleus were producing negative linear polarization. Polarimetric imaging of other comets have previously demonstrated a very strong negative polarization in their innermost coma, up to $P_Q = -6\%$ (Hadamcik and Levasseur-Regourd 2003). This phenomenon is referred in the literature as the *circumnucleus* halo. Therefore, our polarimetric observations suggest the presence of the circumnucleus halo in comet 46P/Wirtanen. This strong negative polarization is evidence of weakly absorbing dust particles in the coma (e.g., Zubko et al., 2012).

While imaging polarimetry of other comets suggests the presence of jet- and/or arch-like structures with strong positive polarization (e.g., Hadamcik and Levasseur-Regourd 2003; Hines et al., 2014), we do not find such features in the 46P/Wirtanen coma. Instead, we detect a nearly isotropic growth of the degree of linear polarization at cometocentric distances in excess of 5,000 km. Polarization in this area appears in good accordance with what was found in jets/arcs; whereas, the isotropic appearance may suggest the existence of numerous jets that merge together. The latter possibility is consistent with the assumption of large active areas on the small 46P/Wirtanen nucleus (e.g., Fink et al., 1998). The outer coma with high positive polarization hardly affects the aperture-averaged polarimetric response (see Table 2). Nevertheless, its presence in 46P/Wirtanen has a significant implication with regard to the small gaseous contamination of the broadband *V* filter measurements. Indeed, the number of $C_2$ molecules tends to increase with the distance from the nucleus and, therefore, their low-polarization response should be more explicitly seen at larger projected distance. Instead, we observe the opposite trend, the positive linear polarization unambiguously grows with distance from the nucleus.

Polarimetric observations of 46P/Wirtanen α ≈ 32.6° − 46.4° may suggest that it belongs to comets with low maximum of positive polarization $P_{max}$ (see Fig. 2; and Chernova et al., 1993; Levasseur-Regourd et al., 1996). Coma in such comets is predominantly populated by weakly absorbing particles (Zubko et al., 2016). On the other hand, a low-$P_{max}$ comet is a rare object (e.g., Zubko et al., 2020). It is worth noting, however, that the polarimetric response in 46P/Wirtanen also closely resembles what was previously detected in comet C/1975 V1 (West) with the broadband *V* filter, but at significantly larger aperture (~80,000 km) (Kiselev and Chernova 1978). While the *V*-filter polarimetry of comet C/1975 V1 (West) was limited to a relatively small phase angle, α < 52°, Kiselev and Chernova (1978) also measured polarization in this comet using a green-continuum filter, but over a much wider range of α. This extended dataset suggests that C/1975 V1 (West) belongs to the high-$P_{max}$ comets. Interestingly, the photometric colour of comet West was found to be either neutral or blue (see discussion of Zubko et al., 2014), which is consistent with our photometric measurements of 46P/Wirtanen. This similarity suggests a likeness of the dust populations in both comets. In order to verify this similarity, one needs to constrain $P_{max}$ in 46P/Wirtanen with greater confidence. This goal could be achieved in future apparitions of the comet.

**Acknowledgment**

I.L. thanks the SAIA Programme for financial support. The work by S.S.K. was supported by the National Research Foundation of Korea funded by the Ministry of Science and ICT (2019R1A2C1009004). The authors thank an anonymous referee for constructive review.

**Data Availability**

The data underlying this article are available in the article.

# References

A'Hearn, M. F. 1982, Comets. University of Arizona Press, Tucson, pp. 433–460.

Boehnhardt, H., Delahodde, C., Sekiguchi, T., et al. 2002, Astron. Astroph. 583, A31.

Bohren, C.F., Huffman, D.R. 1983, Absorption and Scattering of Light by Small Particles. Wiley, New York.

Ciarniello, M., Capaccioni, F., Filacchione, G., et al. 2015, Astron. Astroph. 387, 1107–1113.

Chernova, G.P., Kiselev, N.N., Jockers, K. 1993, Icarus 103, 144–158.

Chornaya, E., Zubko, E., Luk'yanyk, I., et al. 2020a, Icarus 337, 113471.

Chornaya, E., Zakharenko, A.M., Zubko, E., et al. 2020b, Icarus, 350, 113907.

Clairemidi, J., Moreels, G., Krasnopolsky, V. A. 1990, Icarus 86, 115–128.

Cochran, A.L. 1985, Astron. J. 90, 2609–2614.

Combi, M.R., Harris, W.M., Smyth, W. H. 2004, Comets II. University of Arizona Press, Tucson, pp. 523–552.

Dorschner, J., Begemann, B., Henning, T., et al. 1995, Astron. Astrophys. 300, 503–520.

Farnham, T.L., Schleicher, D.G., A'Hearn, M.F. 2000, Icarus 147, 180–204.

Fink, U., Combi, M.R., DiSanti M.A. 1991, Astroph. J. 383, 356–371.

Fink, U., Hicks, M.D., Fevig, R.A., Collins, J. 1998, Astron. Astroph. 335, L37–L45.

Fomenkova, M.N., Kerridge, J.F., Marti, K., McFadden, L.-A. 1992, Science 258, 266–269.

Gehrz, R.D., Ney, E.P. 1992, Icarus 100, 162–186.

Hadamcik, E., Levasseur-Regourd, A.-C. 2003a, J. Quant. Spectrosc. Radiat. Transfer 79-80, 661–678.

Hadamcik, E., Levasseur-Regourd, A.-C. 2003b, Astron. Astroph. 403, 757–768.

Hadamcik, E., Sen, A.K., Levasseur-Regourd, A.-C., et al. 2014, Meteoritics Planet. Sci. 49, 36–44.

Hanner, M. S., Bradley, J. P. 2004, Comets II. University of Arizona Press, Tucson, pp. 555–564.

Hines, D.C., Videen, G., Zubko, E., et al. 2014, Astroph. J. Lett. 780, L32.

Holmberg, J., Flynn, C., Portinari, L. 2006, Mon. Not. Roy. Astron. Soc. 367, 449–453.

Ishii, H.A., Bradley, J.P., Dai, Z.R., et al. 2008, Science 319, 447–450.

Ivanova, A.V., Korsun, P.P., Borisenko, S.A., Ivashchenko, Yu.N. 2013, Sol. System Res. 47, 71–79.

Ivanova, O., Zubko, E., Videen, G., et al. 2017, Mon. Not. Roy. Astron. Soc. 469, 2695–2703.

Jackson, W.M., Bao, Y., Urdahl, R.S. 1991, J. Geophys. Res. 96, 17569–17572.

Jewitt, D.C., Soifer, B.T., Neugebauer, G., Matthews, K., Danielson, G.E. 1982, Astron. J. 87, 1854–1866.

Kelley, M.S.P., Bodewits, D., Ye, Q., et al. 2019, Research Notes of the AAS 3, 126.

Kiselev, N. N., Chernova, G. P. 1978, Sov. Astron. 22, 607–611.

Krasnopol'skii, V.A., Moreels, G., Gogoshev, M., et al. 1986, Sov. Astron. Lett. 12, 259–262.

Kochergin, A., Zubko, E., Husárik, M., et al. 2019, Research Notes of the AAS 3, 152.

Kurchakov, A.V., Nosov, I.V., Rspaev, F.K., Churyumov, K.I. 1986, Komet. Tsir. No. 350, 4 [In Russian].

Kwon, Y.G., Ishiguro, M., Kuroda, D., et al. 2017, Astron. J. 154, 173.

Lamy, P.L., Toth, I., Jorda, L., et al. 1998, Astron. Astroph. 335, L25–L29.

Lara, L.-M., Licandro, J., Oscoz, A., Motta, V. 2003, Astron. Astroph. 399, 763–772.

Le Borgne, J.F., Leroy, J.L., Arnaud, J. 1987, Astron. Astrophys. 187, 526–530.

Levasseur-Regourd, A.-C., Hadamcik, E., Renard, J. B. 1996, Astron. Astroph. 313, 327–333.

Levasseur-Regourd, A.-C., Hadamcik, E., Lasue, J. 2005, Highlights of Astronomy 13, 498–500.

Luk'yanyk, I.V., Zubko, E., Husárik, M., et al. 2019, Mon. Not. Roy. Astron. Soc. 485, 4013–4023.

Meech, K.J., Bauer, J.M., Hainaut, O.R. 1997, Astron. Astroph. 326, 1268–1276.

Picazzio, E., Luk'yanyk, I.V., Ivanova, O.V., et al. 2019, Icarus 319, 58–67.

Rousselot, P., Clairemidi, J., Moreels, G. 1994, Astron. Astrophys. 286, 645–653.

Sen, A.K., Joshi, U.C., Deshpande, M.R. 1989, Astron. Astrophys. 217, 307–310.

Sen, A.K., Hadamcik, E., Botet, R., et al. 2019. Mon. Not. Roy. Astron. Soc. 487, 4809–4818.

Schulz, R., Arpigny, C., Manfroid, J., et al. 1998, Astron. Astroph. 335, L46–L49.

Weiler, M., Rauer, H., Knollenberg, J., et al. 2003, Astron. Astroph. 403, 313–322.

Wooden, D.H., Ishii, H.A., & Zolensky, M.E. 2017, Philosophical Transactions of the Royal Society A 375, id. 20160260.

Zubko, E., Chornaya, E. 2019, Research Notes of the AAS 3, 45.

Zubko, E., Chornaya, E., Zheltobryukhov, M., et al. 2020, Icarus 336, 113453.

Zubko, E., Muinonen, K., Shkuratov, Yu., et al. 2012, Astron. Astrophys., 544, L8.

Zubko, E., Muinonen, K., Shkuratov, Yu., Videen, G. 2013, Mon. Not. Roy. Astron. Soc. 430, 1118–1124.

Zubko, E., Muinonen, K., Videen, G., Kiselev, N. 2014, Mon. Not. Roy. Astron. Soc. 440, 2928–2943.

Zubko, E., Videen, G., Hines, D.C., et al. 2015, Planet. Space Sci. 118, 138–163.

Zubko, E., Videen, G., Hines, D.C., Shkuratov, Yu. 2016, Planet. Space Sci. 123, 63–76.

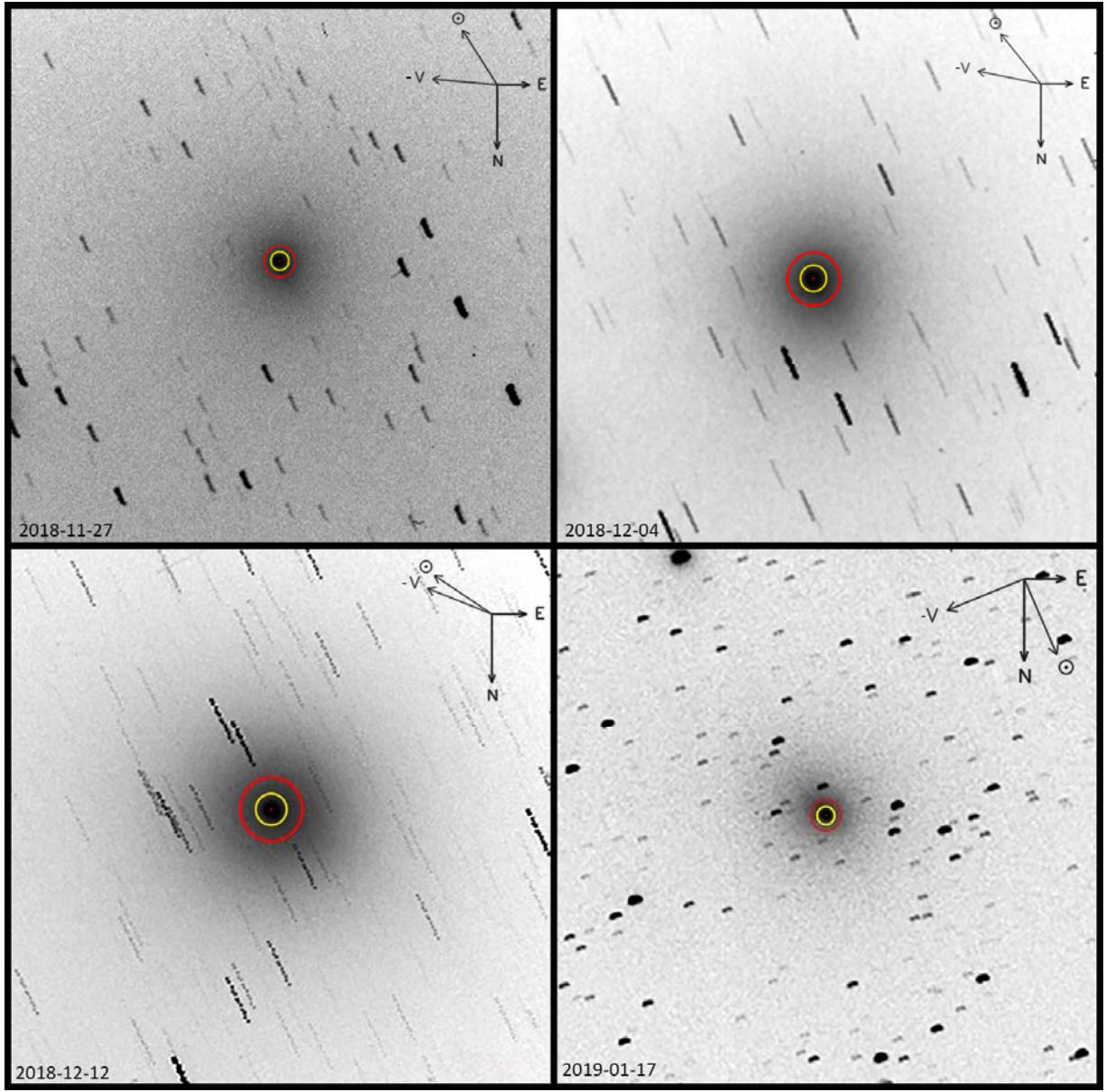

Fig. 1: Images of comet 46P/Wirtanen taken through the *V* filter on four epochs embracing the period of our monitoring of its polarization at the Ussuriysk Astrophysical Observatory. Two circles demonstrate apertures with radii of 5,000 km and 10,000 km centered upon the nucleus.

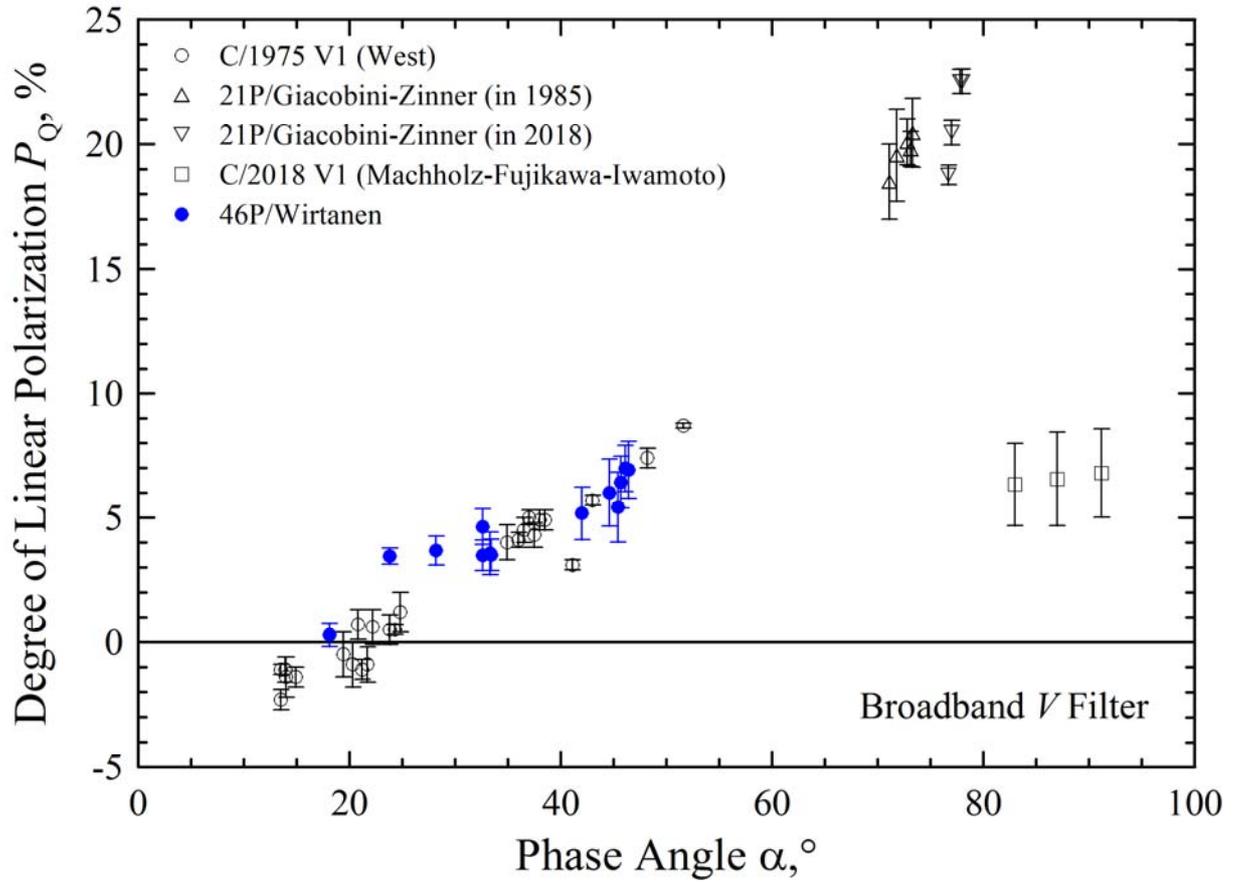

Fig. 2: Degree of linear polarization $P_Q$ as a function of phase angle α of comet 46P/Wirtanen in its 2018 apparition (blue points). For comparison, we also show polarization in three other comets measured with the broadband *V* filter (open symbols); see text for reference to sources of these observations.

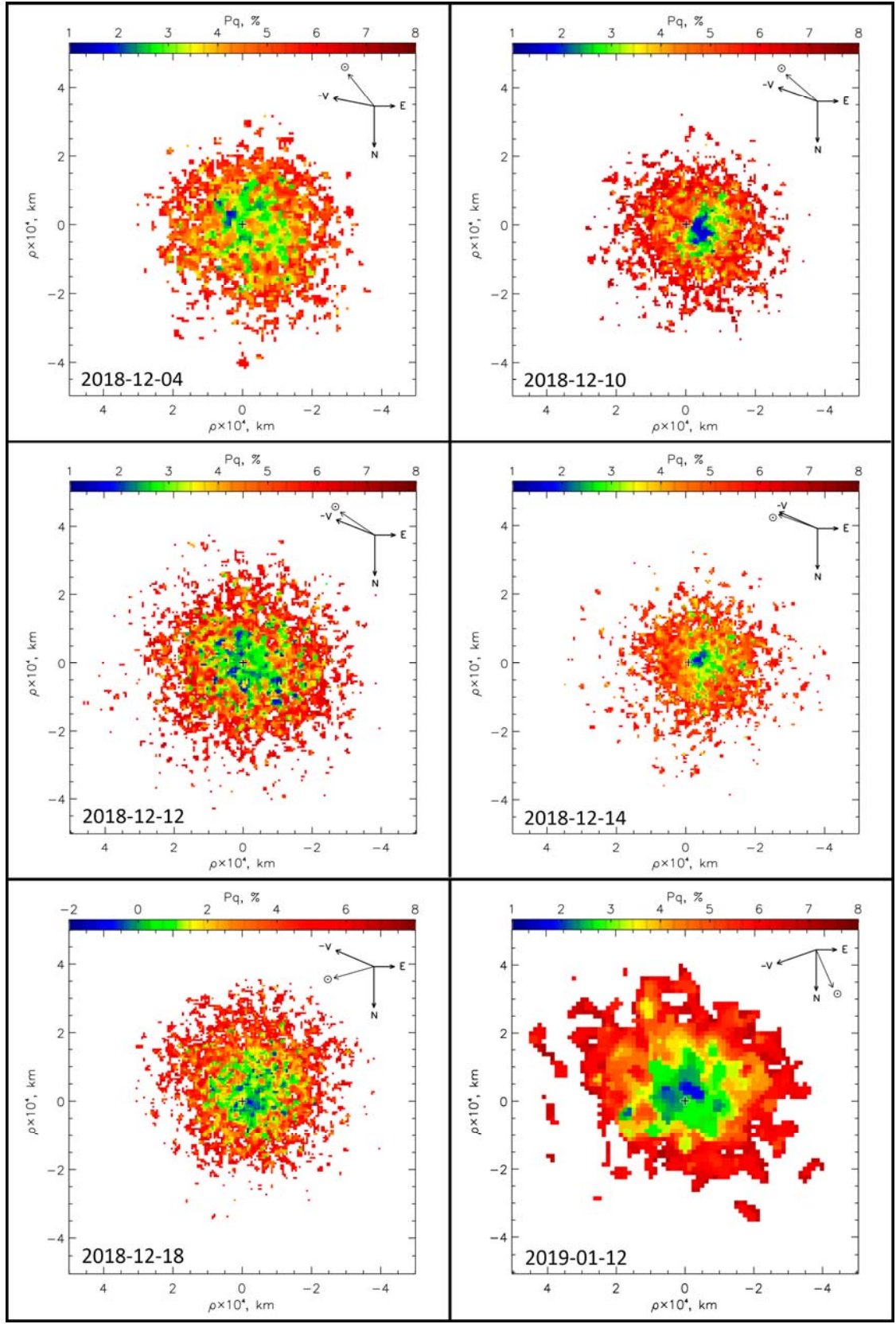

Fig. 3: Polarimetric images of comet 46P/Wirtanen during six epochs. The nucleus is located at the origin of coordinates (0,0).

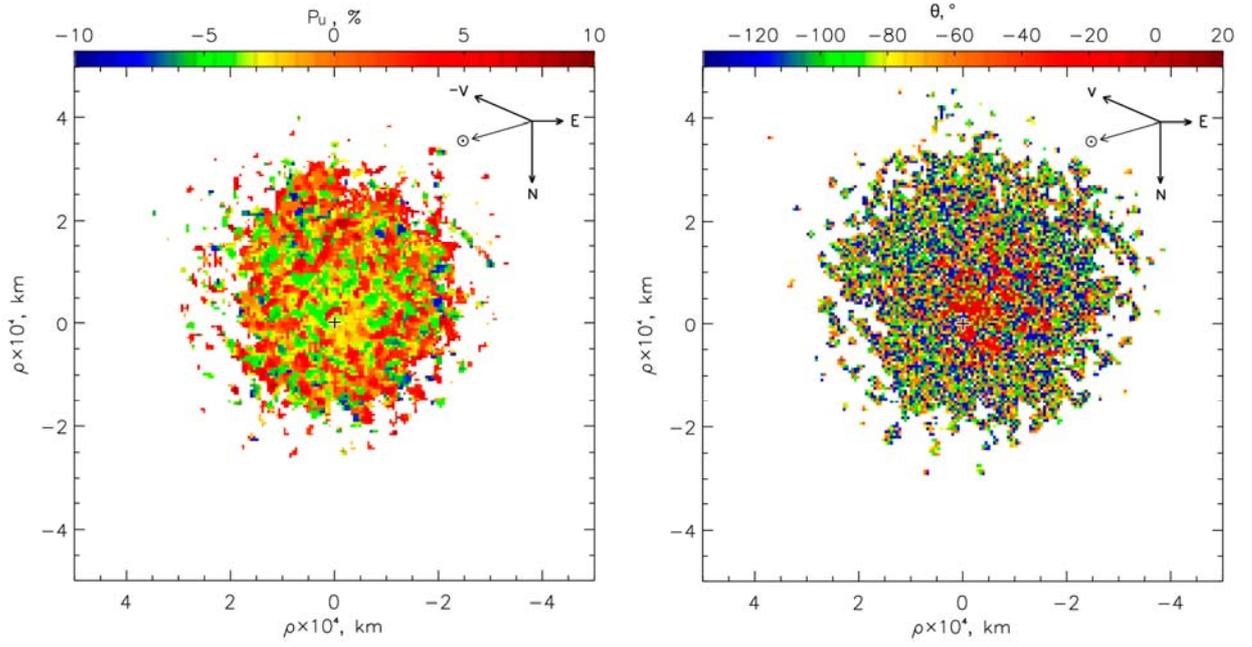

Fig. 4: Asymmetry in spatial distribution of the polarization $P_U$ (left) and angle θ (right) in comet 46P/Wirtanen on December 18 of 2018, at phase angle α = 18.1°. The nucleus is located at the origin of coordinates (0,0).

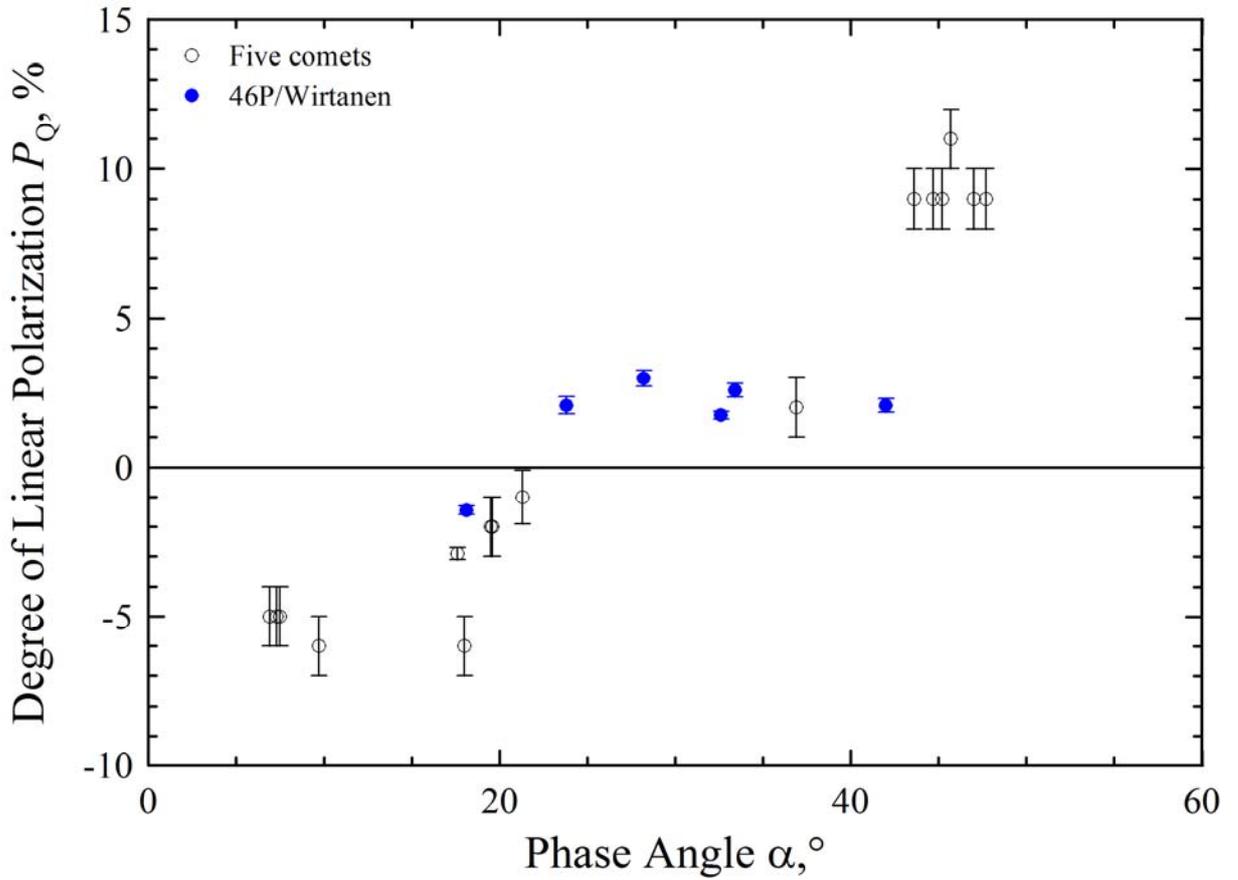

Fig. 5: Minimum values of the degree of linear polarization $P_Q$ detected in the innermost coma of 46P/Wirtanen during the six epochs shown in Fig. 3. Also shown are polarizations detected in the innermost coma of five other comets displaying circumnucleus halo features (data adapted from Hadamcik and Levasseur-Regourd 2003a; 2003b).

**Tables**

Table 1: Log of observations of comet 46P/Wirtanen in the 2018–2019 apparition.

| | Date (UTC) | Numbers of exposures | Exposure (sec) | Filter | r (au) | Δ (au) | α (°) | Telescope | Scale (km/px) | Seeing (FWHM in px) | Airmass | Magnitude | | | |
|---|---|---|---|---|---|---|---|---|---|---|---|---|---|---|---|
| | | | | | | | | | | | | B | V | R | I |
| | | | | | | | | **5000 km** | | | | | | | |
| ▲ | 2018-11-16.508 | 36 | 40 | V | 1.114 | 0.191 | 45.4 | ORI-22 | 1300 | 2.23 | 4.677 | - | 11.23±0.13 | - | - |
| ▲ | 2018-11-27.481 | 36 | 30 | V | 1.076 | 0.137 | 46.4 | ORI-22 | 933 | 2.15 | 3.125 | - | 10.04±0.06 | - | - |
| ▲ | 2018-11-28.535 | 36 | 30 | V | 1.073 | 0.132 | 46.1 | ORI-22 | 900 | 2.22 | 2.498 | - | 9.74±0.06 | - | - |
| ▲ | 2018-11-29.471 | 36 | 30 | V | 1.071 | 0.127 | 45.7 | ORI-22 | 865 | 2.14 | 2.793 | - | 9.66±0.11 | - | - |
| ▲ | 2018-12-01.491 | 36 | 30 | V | 1.067 | 0.118 | 44.6 | ORI-22 | 803 | 2.10 | 2.492 | - | 9.44±0.07 | - | - |
| ▲ | 2018-12-04.446 | 36 | 30 | V | 1.062 | 0.106 | 42.0 | ORI-22 | 722 | 1.99 | 2.300 | - | 9.41±0.05 | - | - |
| ▲ | 2018-12-10.451 | 36 | 30 | V | 1.056 | 0.086 | 32.6 | ORI-22 | 586 | 2.17 | 1.613 | - | 8.58±0.08 | - | - |
| ▲ | 2018-12-12.458 | 36 | 30 | V | 1.055 | 0.081 | 28.2 | ORI-22 | 552 | 2.16 | 1.461 | - | 8.25±0.17 | - | - |
| ▼ | 2018-12-14.554 | 36 | 30 | V | 1.056 | 0.078 | 23.8 | ORI-22 | 531 | 2.20 | 1.141 | - | 8.45±0.07 | - | - |
| ▼ | 2018-12-18.768 | 36 | 30 | V | 1.058 | 0.079 | 18.1 | ORI-22 | 538 | 2.16 | 1.803 | - | 8.52±0.12 | - | - |
| ▼ | 2019-01-12.495 | 30 | 30 | V | 1.133 | 0.185 | 33.4 | ORI-22 | 1260 | 2.19 | 1.339 | - | 11.17±0.10 | - | - |
| ▼ | 2019-01-13.596 | 36 | 30 | V | 1.138 | 0.191 | 33.3 | ORI-22 | 1300 | 2.21 | 1.088 | - | 11.19±0.11 | - | - |
| ▼ | 2019-01-17.810 | 36 | 30 | V | 1.160 | 0.217 | 32.6 | ORI-22 | 1477 | 2.22 | 1.177 | - | 11.52±0.09 | - | - |
| ▼ | 2019-02-08.503 | 40 | 60 | B,V,R,I | 1.300 | 0.370 | 27.6 | ChV-500 | 306 | 2.75 | 1.223 | 14.46±0.10 | 13.81±0.03 | 13.62±0.03 | 13.43±0.04 |
| ▼ | 2019-02-10.488 | 32 | 60 | B,V,R,I | 1.315 | 0.387 | 27.2 | ChV-500 | 320 | 2.80 | 1.209 | 15.36±0.11 | 14.63±0.08 | 14.35±0.07 | 14.15±0.06 |

▲ – prior perihelion passage, ▼ – after perihelion passage

Table 2. Polarimetric response in comet 46P/Wirtanen.

| | Date, UTC | α,° | P, % | θ, ° | $P_Q$, % |
|---|---|---|---|---|---|
| | *5000 km* | | | | |
| ▲ | 2018-11-16.508 | 45.4 | 5.43 ± 1.22 | 92.07 ± 3.34 | 5.42 ± 1.40 |
| ▲ | 2018-11-27.481 | 46.4 | 6.92 ± 1.00 | 90.79 ± 3.36 | 6.92 ± 1.15 |
| ▲ | 2018-11-28.535 | 46.1 | 7.03 ± 0.81 | 93.62 ± 3.44 | 6.98 ± 0.93 |
| ▲ | 2018-11-29.471 | 45.7 | 6.44 ± 0.92 | 96.00 ± 3.47 | 6.43 ± 1.05 |
| ▲ | 2018-12-01.491 | 44.6 | 6.06 ± 1.18 | 93.80 ± 4.15 | 6.01 ± 1.35 |
| ▲ | 2018-12-04.446 | 42.0 | 5.26 ± 0.93 | 95.36 ± 2.62 | 5.17 ± 1.06 |
| ▲ | 2018-12-10.451 | 32.6 | 4.67 ± 0.61 | 86.19 ± 2.73 | 4.63 ± 0.72 |
| ▲ | 2018-12-12.458 | 28.2 | 3.75 ± 0.50 | 84.43 ± 5.36 | 3.67 ± 0.58 |
| ▼ | 2018-12-14.554 | 23.8 | 3.46 ± 0.28 | 87.88 ± 3.59 | 3.45 ± 0.33 |
| ▼ | 2018-12-18.768 | 18.1 | 3.06 ± 0.40 | -47.66 ± 3.75 | 0.28 ± 0.47 |
| ▼ | 2019-01-12.495 | 33.4 | 3.66 ± 0.57 | 98.68 ± 1.95 | 3.50 ± 0.63 |
| ▼ | 2019-01-13.596 | 33.3 | 3.57 ± 0.75 | 88.74 ± 2.55 | 3.56 ± 0.86 |
| ▼ | 2019-01-17.810 | 32.6 | 3.50 ± 0.53 | 92.90 ± 2.66 | 3.48 ± 0.61 |
| | *10000 km* | | | | |
| ▲ | 2018-11-16.508 | 45.4 | 6.98 ± 0.95 | 89.30 ± 6.63 | 6.98 ± 1.07 |
| ▲ | 2018-11-27.481 | 46.4 | 7.15 ± 0.71 | 90.52 ± 3.25 | 7.15 ± 0.82 |
| ▲ | 2018-11-28.535 | 46.1 | 6.70 ± 0.41 | 95.06 ± 2.53 | 6.60 ± 0.47 |
| ▲ | 2018-11-29.471 | 45.7 | 6.34 ± 0.99 | 88.17 ± 4.20 | 6.33 ± 1.13 |
| ▲ | 2018-12-01.491 | 44.6 | 6.57 ± 0.82 | 90.43 ± 2.97 | 6.57 ± 0.94 |
| ▲ | 2018-12-04.446 | 42.0 | 5.91 ± 0.73 | 85.62 ± 2.28 | 5.85 ± 0.83 |
| ▲ | 2018-12-10.451 | 32.6 | 4.30 ± 0.43 | 87.06 ± 3.44 | 4.27 ± 0.49 |
| ▲ | 2018-12-12.458 | 28.2 | 3.70 ± 0.39 | 91.57 ± 4.35 | 3.69 ± 0.45 |
| ▼ | 2018-12-14.554 | 23.8 | 3.59 ± 0.38 | 89.31 ± 2.56 | 3.59 ± 0.44 |
| ▼ | 2018-12-18.768 | 18.1 | 3.57 ± 0.65 | -49.09 ± 2.05 | 0.51 ± 0.50 |
| ▼ | 2019-01-12.495 | 33.4 | 3.77 ± 0.57 | 90.92 ± 4.30 | 3.77 ± 0.65 |
| ▼ | 2019-01-13.596 | 33.3 | 3.99 ± 0.66 | 82.02 ± 4.80 | 3.84 ± 0.76 |
| ▼ | 2019-01-17.810 | 32.6 | 4.10 ± 0.51 | 94.51 ± 2.94 | 4.05 ± 0.60 |

▲ – prior perihelion passage, ▼ – after perihelion passage

Table 3. Colour indices and their corresponding colour slopes *S′* measured in comet 46P/Wirtanen measured with aperture having radius of 5,000 km.

| Date, UTC | α,° | *B–V*, mag | *V–R*, mag | *R–I*, mag | *S′*$_{B-V}$, % per 0.1 μm | *S′*$_{V-R}$, % per 0.1 μm | *S′*$_{R-I}$, % per 0.1 μm |
|---|---|---|---|---|---|---|---|
| 2019-02-08.503 | 27.6 | 0.01 ± 0.15 | –0.16 ± 0.07 | –0.14 ± 0.08 | 0.6 ± 11.5 | –16.7 ± 7.1 | –8.7 ± 4.8 |
| 2019-02-10.488 | 27.2 | 0.09 ± 0.21 | –0.07 ± 0.16 | –0.13 ± 0.14 | 6.9 ± 16.1 | –7.5 ± 16.3 | –8.1 ± 8.4 |